\documentclass[aip,rsi, amsmath,amssymb,floatfix,reprint]{revtex4-2}
\usepackage{graphicx}% Include figure files
\usepackage{dcolumn}% Align table columns on decimal point
\usepackage{bm}% bold math

\begin{document}

\title{Emittance self-compensation in blow-out mode}

\author{Georgii Shamuilov}
\affiliation{FREIA Laboratory, Uppsala University, L\"agerhyddsv\"agen 1, Uppsala 75120, Sweden}

\author{Anatoliy Opanasenko}
\affiliation{NSC/KIPT, Akademicheskaya 1, Kharkiv 61108, Ukraine}

\author{K\'evin Pepitone}
\affiliation{FREIA Laboratory, Uppsala University, L\"agerhyddsv\"agen 1, Uppsala 75120, Sweden}

\author{Zolt\'an Tibai}
\affiliation{Institute of Physics, University of P\'ecs, P\'ecs 7624, Hungary}

\author{Vitaliy Goryashko}
\email[e-mail: ]{vitaliy.goryashko@physics.uu.se}
\affiliation{FREIA Laboratory, Uppsala University, L\"agerhyddsv\"agen 1, Uppsala 75120, Sweden}

\date{\today}

\begin{abstract}
We report an unusual regime of emittance self-compensation in an electron bunch generated in blow-out mode by a radio-frequency photocathode gun.
Simulations clearly show an initial growth and a subsequent self-compensation of projected emittance in a divergent electron bunch originating from the effects of: (i) strong space-charge forces of mirror charges on the cathode, (ii) an energy chirp in the bunch and (iii) substantial re-shaping of the electron bunch. 
Furthermore, we show analytically and numerically how a complex interplay between these effects leads to emittance self-compensation in free space -- the effect that is normally observed only in a focusing magnetic field.  
\end{abstract}

% In contrary to standard compensation of effective emittance in a convergent electron bunch by means of an external field of a solenoid,


\maketitle

%High-brightness electron beams are crucial 
%for ultrafast electron diffraction~(UED) experiments 
%and accelerator-based light sources~\cite{Sciaini2011,Miller2014,Emma2010,FERRARIO2000,Hastings2006,Li2009,Murooka2011,Chatelain2012,Giret2013,Weathersby2015}. 
%The brightness of an electron beam can  generally be defined 
%as the number of particles per unit volume 
%of the total phase space~\cite{Bazarov2009}. 
%Thus, a higher brightness requires \textit{more} particles within a \textit{smaller} phase space volume.
%A larger number of particles in the beam implies a higher electric charge density. The Coulomb interaction between the beam particles then constitutes stronger space-charge fields. These fields impose a limitation on the {\it maximum current density} of particles that leave the cathode. 

The space-charge force in an ellipsoidal electron bunch with a uniform density distribution
changes linearly in the bunch in any direction. Because of this nature of the space-charge force, 
there is no emittance growth associated with it. Such a bunch is an ideal object in accelerator physics~\cite{Reiser2008}. 
J.~Luiten and co-workers~\cite{Luiten2004} proposed a simple method for 
the formation of ellipsoidal electron bunches with a uniform density distribution.
In the method, demonstrated experimentally by P.~Musumeci and co-workers~\cite{Musumeci2008}, 
an initially short, pancake-like bunch expands to a fully-fledged 
uniformly charged ellipsoidal bunch thanks to its space-charge force, similarly to blowing out soap bubbles. 
Hence, the method is named as blow-out mode. 
The initial density distribution in the bunch must have a half-circular profile 
in the radial direction whereas in the longitudinal direction 
the distribution can be arbitrary as long as the bunch is short enough.
Mathematically, the required initial bunch distribution has the form $f(s)\sqrt{1-r^2/R^2}$,
where $f(s)$ is the distribution along the longitudinal bunch coordinate $s$,
and $\sqrt{1-r^2/R^2}$ describes the transverse distribution with $R$ being the \textit{centre-to-edge} distance. 

Furthermore, J.~Luiten and co-workers found that the desired uniform ellipsoidal distribution is formed
when the accelerating field $E_{acc}$ is much larger than 
the space-charge field of the image charge induced on the cathode,
i.e. $|E_{acc}| \gg |\sigma|/\epsilon_0$. Here, $\sigma$ is the surface charge density of the bunch and 
$\epsilon_0$ is vacuum permittivity.  

However, the goal of our study was to find the conditions for the maximum 4D brightness in blow-out mode, 
which implies the minimum normalised transverse emittance. Our extensive numerical simulations show that 
the \textit{lowest emittance for a given charge} is achieved when $E_\mathrm{acc} \approx 0.35~\sigma/\epsilon_0$. 
Furthermore, we found that in this regime of a strong space-charge force,
natural self-compensation of bunch emittance occurs. The unusual feature of this 
emittance self-compensation is that it occurs in a divergent electron bunch,
which is opposite to what is known about classical emittance compensation~\cite{Carlsten1989,Carlsten1995,Serafini1997,Rosenzweig2006}. 
Note that the desired uniform ellipsoidal distribution can be simultaneously achieved. 

\begin{figure}[tb]
\centering
\includegraphics[width=\linewidth]{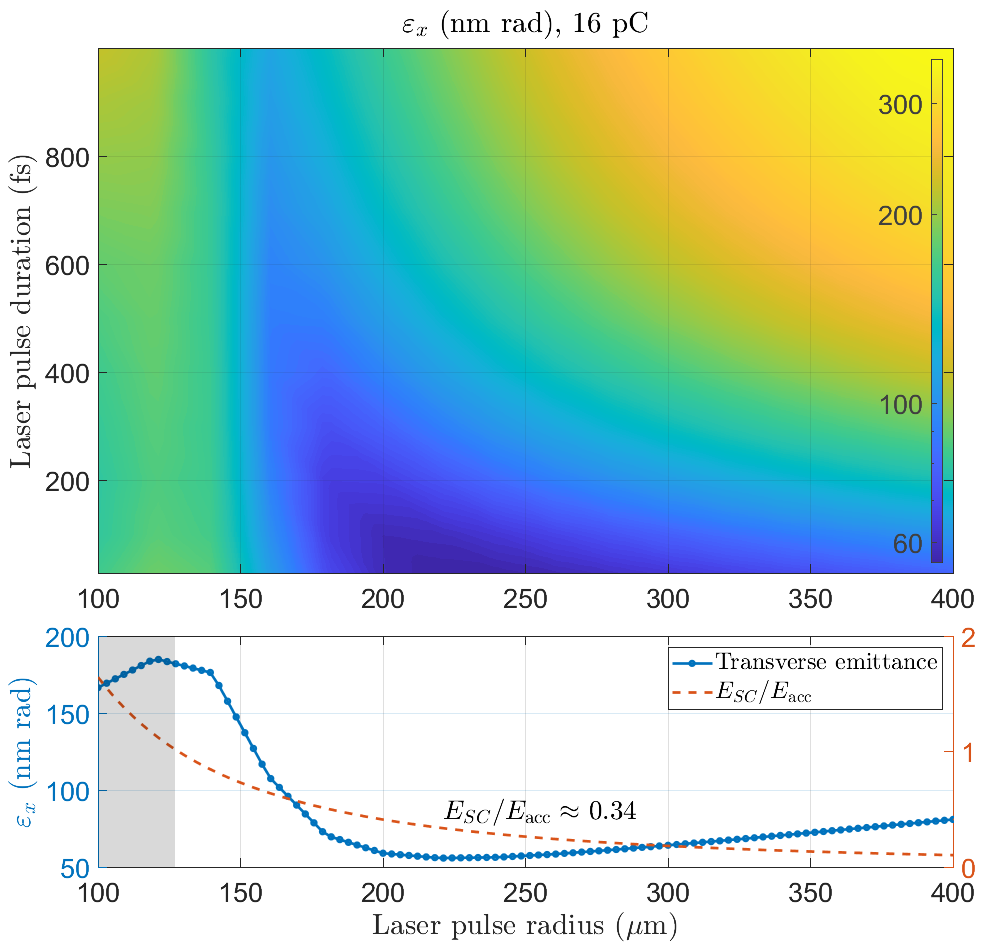}
\caption{Upper plot: colour map of emittance as a function of the initial \textit{rms} bunch duration (changes from 30~fs to 1~ps) and bunch radius $R$ (note that $R$ is the \textit{centre-to-edge distance}!). Bottom plot: emittance (left ordinate axis) and $E_\mathrm{SC}/E_\mathrm{acc}$ (right ordinate axis) as a function of $R$ for the initial rms duration of 30~fs. The space-charge field on the cathode is estimated as $E_\mathrm{SC} = \sigma/\epsilon_0$. For $R<130\;\mu$m, the shaded area on the left, less than 16~pC of charge is extracted because of the formation of the virtual cathode due to the strong longitudinal space-charge force $eE_\mathrm{SC}$. }
\label{fig:emittance_map}
\end{figure}

Figure~\ref{fig:emittance_map} reports the results of massive simulations 
for the emittance of an electron beam generated in blow-out mode in an APEX-like~\cite{Wells2016} continous wave radio-frequency gun
operated at 352 MHz with a peak accelerating field of 35 MV/m. Note that the results were independently cross-checked using three different codes: ASTRA, GPT and RFtrack. The discrepancy between the results is less than 10\% and below the results from ASTRA simulations are shown. 
In the simulations, starting from the Fermi-Dirac distribution of electrons in a copper cathode,  
electron bunches are ejected from the cathode via photo-emission using the model~\cite{Dowell2009}. 
The initial electron bunch distribution has the form required for the blow-out mode
$\exp(-t^2/2\sigma_t^2)~\sqrt{1-r^2/R^2}$, where $\sigma_t$ is the initial rms duration of the electron bunch. % and $R$ is the \textit{centre-to-edge} distance.
From the Fermi-Dirac distribution,
the calculated thermal emittance per unit length is 0.4 mm$\cdot$mrad/mm -- the value demonstrated experimentally~\cite{Hauri2010}. After the emission, the electron bunches move solely in the RF field of the gun,  no solenoid is present. 
The laser pulse energy allows for extraction of 16~pC charge if the space-charge field is neglected. The initial electron bunch radius and duration are scanned in a wide range of parameters, even for such small radii that the electron emission is suppressed by the space-charge force. 
The minimum critical radius allowing for full extraction of 16~pC corresponds to around 130~$\mu$m.
In this critical regime, the accelerating field is fully screened by the space-charge field. 

Figure~\ref{fig:emittance_map} shows a clear minimum of the emittance
for the bunch duration of 30~fs (shortest duration in the simulations) and the bunch radius of around 220~$\mu$m.
The resulting lowest emittance $\varepsilon_\mathrm{min}$ for the 16~pC extracted bunch is 25\% above the thermal level.  In this optimum scenario, the longitudinal space-charge force (accounting for the image charge on the cathode) is around 35\% of the accelerating field, which is an unexpectedly high ratio.  Importantly, for a given charge the obtained $\varepsilon_\mathrm{min}$ is around 40\% smaller than the emittance obtained in the blow-out mode of a weak space-charge force ($E_\mathrm{acc} \approx 10~\sigma/\epsilon_0$). Note that in each simulation run, the emittance is recorded as a function of distance from the cathode, and then, the smallest value is plotted in Fig.~\ref{fig:emittance_map}. Hence, in each simulation run, the optimum distance from the cathode is somewhat different but stays approximately 80~mm.  

\begin{figure*}[t]
    \centering
    \includegraphics[width=\linewidth]{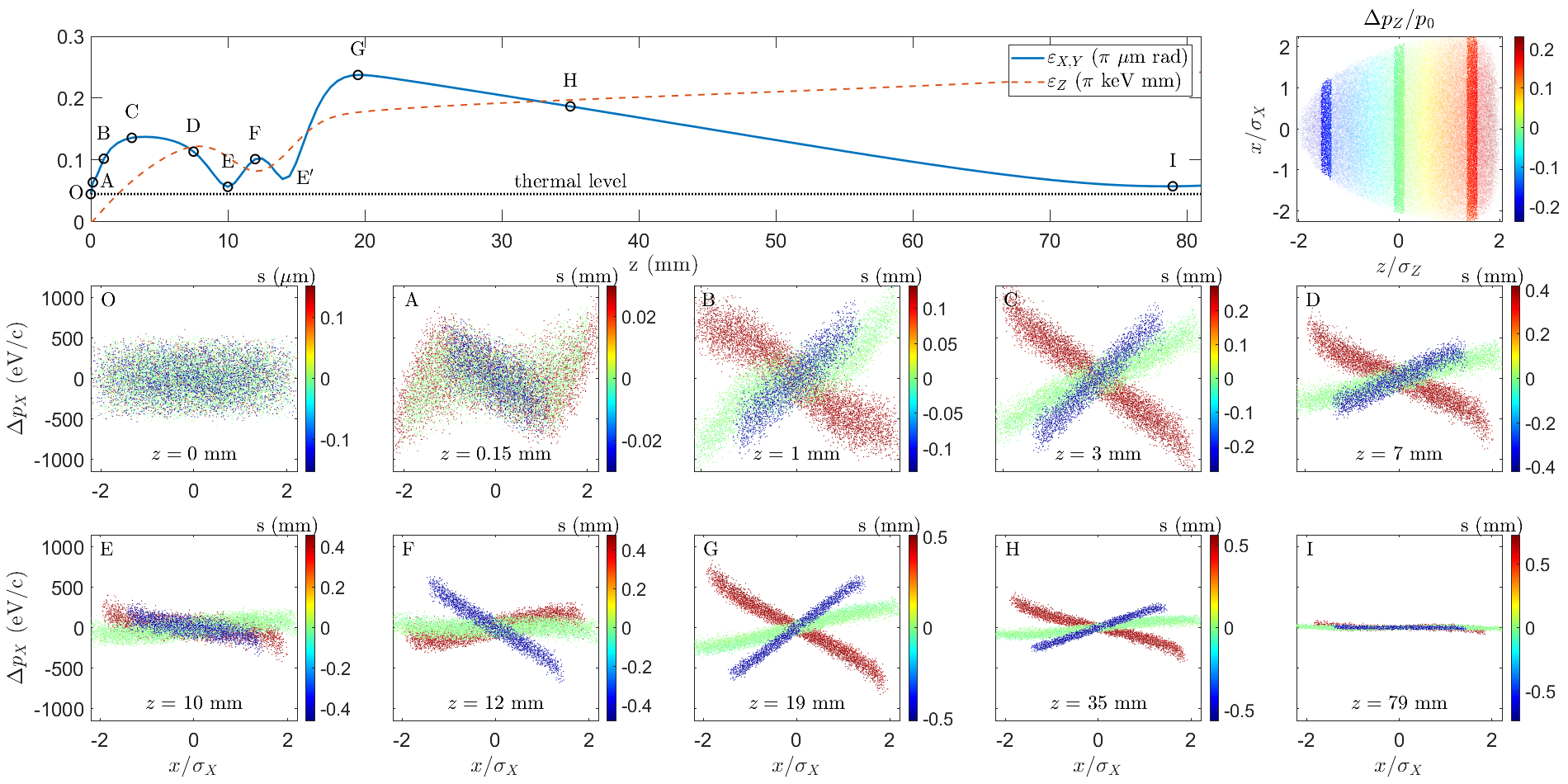}
    \caption{Upper left plot: transverse and longitudinal rms normalised emittance as a function of distance from the cathode. 
    Upper right plot: bunch density distribution projected onto  the $xz$-plane at $z=0.2$~mm. The colour coding shows the relative longitudinal momentum. 
    The three slices used in the analysis of the transverse phase space are highlighted by means of a more intense hue of the colour.
    The second and third rows show the $x$-$p_x$ phase space of the three slices. The linear $x$-$p_x$ correlation calculated for all electrons in the bunch was removed to reveal the fine structure in the phase space. 
    The abscissa coordinate is normalised to the corresponding rms bunch size, 
    which is calculated for each longitudinal distance from the cathode. 
    Note that the phase space distribution at the position E$'$ is visually identical to that in the position E.
    Hence, it is not shown.}
    \label{fig:emittance_evolution}
\end{figure*}

The upper plot in Fig.~\ref{fig:emittance_evolution} shows 
the evolution of emittance as a function of distance from the cathode. After several oscillations, 
the emittance naturally attains its lowest value at 79~mm away from the cathode. 
Figure~\ref{fig:emittance_evolution} also shows a series of distributions of the transverse phase space
for three slices of electrons located in the tail (blue), centre (green) and head (read) of the bunch,
see the plot in the top right corner illustrating the slices used in the analysis hereinafter.
We limit the analysis to the $x$-$p_x$ plane thanks to the rotational symmetry of the problem. 
Note that in all plots \textit{the linear correlation between $x$ and $p_x$ -- calculated for all electrons in the bunch --
% it actually doesn't matter, I checked. Phase space portraits look exactly the same when correlation is removed only for slices as for the entire bunch
is removed} to reveal the fine structure of the $x$-$p_x$ distribution. 
Also, in all plots the \textit{x}-coordinate is shown in the units of the rms beam size $\sigma_x$ calculated at each $z$ position.  

\begin{figure}[tb]
    \centering
    \includegraphics[width=0.939\linewidth]{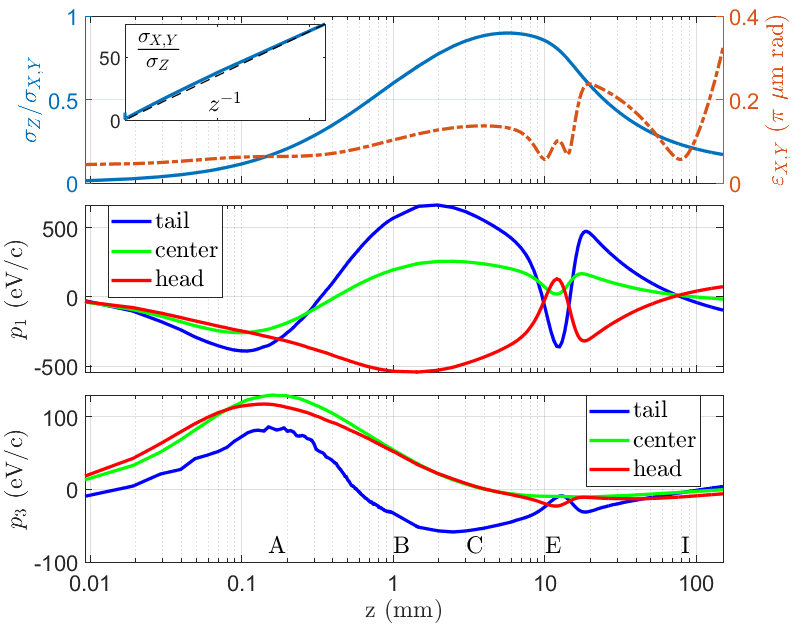}
    \caption{Upper plot: the ratio of the longitudinal to transverse bunch size and bunch emittance vs distance from the cathode. Inset demonstrates the linear $z^{-1}$ dependence of the inverse aspect ratio.
    Middle and bottom plots: the coefficients of the linear and cubic correlations between $x$ and $p_x$. Labels correspond to the phase space portraits in Fig.~\ref{fig:emittance_evolution}.}
    \label{fig:slices_fit_coeffs}
\end{figure}

The details of the bunch dynamics are further analysed in Fig.~\ref{fig:slices_fit_coeffs} that reports: (i) the aspect ratio of the longitudinal to transverse bunch length,
(ii) the coefficients of the linear and cubic correlations between $x$ and $p_x$ for the three slices mentioned above.  These coefficients are the result of a least-squares fit having the form $p_x = p_0 + p_1(x/\sigma_x) + p_3(x/\sigma_x)^3$.   

\begin{figure}[tb]
    \centering
    \includegraphics[width=\linewidth]{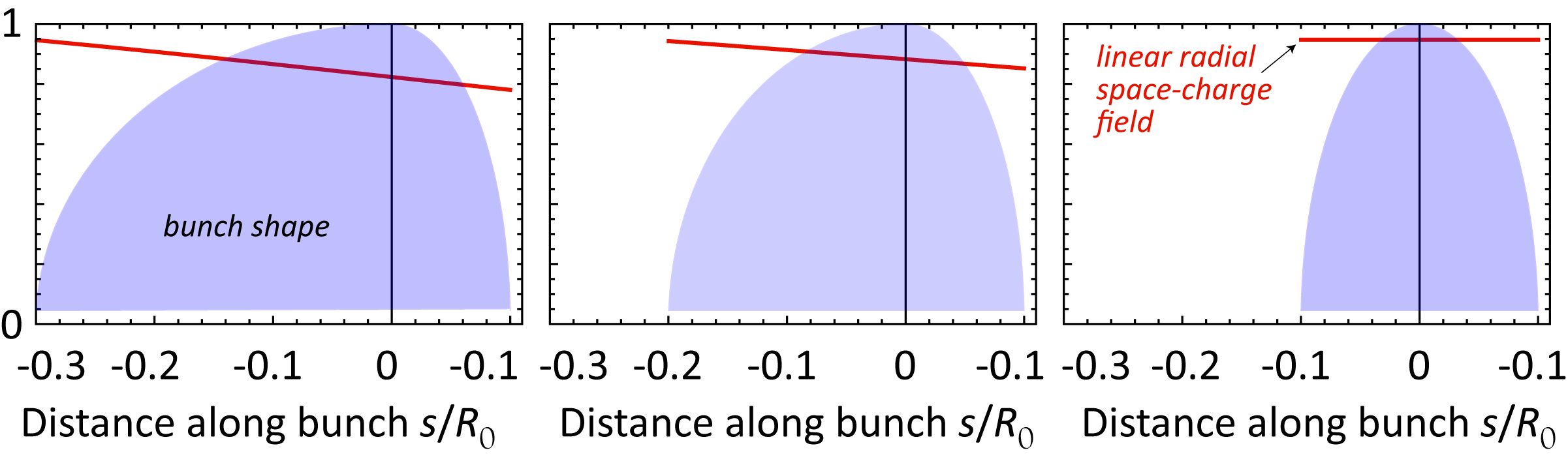}
    \caption{Linear component of the radial space-charge field for different uniform distributions of the electron bunch density.
    The true egg-like bunch is modelled by the sum of halves of two ellipsoidal bunches with different axial semi-axes.}
    \label{fig:space-charge_field}
\end{figure}

Let us examine in detail the evolution of emittance in Fig.~\ref{fig:emittance_evolution} and we start with considering the region OA. In this region, the total space-charge field is given by the superposition of the space-charge field of the bunch itself and its mirror charge on the cathode. The radial component of the space-charge field reads
\begin{equation}
    E_r = - \frac{\partial}{\partial r} (\varphi_- + \varphi_+), \quad
    \varphi_\pm = \frac{1}{4\pi\epsilon_0} \int \frac{\varrho_\pm dV'}{|\vec r - {\vec r\,}'|},
\end{equation}
where $\varrho_\pm = \pm \sigma \sqrt{1-r^2/R^2} \delta(z \pm z_b)$ is the initial bunch density.
The bunch is assumed to be infinitesimally thin and located at $z_b$. 
By Taylor expanding $E_r$ into the powers of the radial distance $r$,
one obtains the linear and cubic component of $E_r$.

Let us start from the \textit{cubic term} having the form
\begin{equation}
    E_r^{\mathrm{cubic,OA}} = \frac{\sigma}{2\epsilon_0}\, \frac{\zeta_b \rho^3}{(1+4\zeta_b^2)^3},
\end{equation}
where $\zeta_b = z_b/R$ and $\rho = r/R$ are the normalised longitudinal bunch position and radial coordinate, respectively. The cubic non-linearity of $E_r$ gives rise to an emittance increase in the region OA, see plot A in Fig.~\ref{fig:emittance_evolution} and  also the plot of $p_3$ in Fig.~\ref{fig:slices_fit_coeffs}. The maximum of $E_r^{\mathrm{cubic,OA}}$ is reached for $z_b^\mathrm{cr} = 1/(2\sqrt{5})R \approx 0.2 R$. For our simulation parameters, $z_b^\mathrm{cr} \approx 50 \, \mu$m and at this position the gradient of the fit coefficient $p_3$ attains its maximum value. Recall that the shape of the transverse phase space is the result of the cumulative action of $E_r$ (the radial component of the accelerating field is uniform in the gun region except for the output aperture). Hence, $dp_3/dz \propto E_r^\mathrm{cubic}/v_z$, where $v_z$ is the longitudinal velocity.

In the region AB, two effects occur: (i) the transverse phase-space self-linearises and 
(ii) emittance grows because of relative rotation (misalignment) of the slices due to different magnitudes of the linear space-charge force along the bunch. Namely, for $z_b>R/2$ the radial space-charge field $E_r$ is mostly due to the field of the bunch itself and the cubic component might be \textit{estimated} as
\begin{equation}
    E_r^{\mathrm{cubic,AB}} = -\frac{\sigma}{\epsilon_0}\, \zeta_l^2 \rho^3,
\end{equation}
where $\zeta_l = \sigma_b/R$ is the normalised bunch length. The bunch is not infinitesimally thin anymore because of its expansion. Note that now $E_r^{\mathrm{cubic,AB}}$ has a different sign compared to that in the region OA and counteracts the effect of the phase-space non-linearity induced before. From Fig.~\ref{fig:slices_fit_coeffs}, one can see that the non-linear component 
of the phase space distribution almost completely disappears after a few mm from the cathode
thanks to the compensation effect. Appreciate almost linear phase space distributions in the plots A-I of Fig.~\ref{fig:emittance_evolution}.

Now, we turn our attention to the relative motion of electron slices due to the linear component of $E_r$.
First, very near the cathode, $z_b \ll R$, 
all electrons experience the same linear space-charge field
$E_r^\mathrm{lin}=(\sigma/\epsilon_0) (z_br/R^2)$ and  
 all slices get the same transverse kick as it can also be seen from the plot of $p_1$ in Fig.~\ref{fig:slices_fit_coeffs}.
Note that the image charge gives a focusing effect and 
$E_r^\mathrm{lin} = 0$ for the bunch being just on the cathode, $z_b = 0$. 

As the bunch flies away and expands in the region OA, 
it takes an egg-like shape, see the top right plot in Fig.~\ref{fig:emittance_evolution}, 
primarily because of the pulling effect of the longitudinal space-charge force of the image charge on the cathode. 
There is also a weak focusing force in the radial direction due to the image charge.
For the tail, this focusing is stronger than for the head simply because of the shorter distance to the cathode.
In addition, a head-tail energy correlation -- called energy chirp -- 
develops due to the longitudinal space-charge force. 

In the region AC, $z>R/2$, the transverse bunch dynamics is dominated by its own linear space-charge field 
and different slices acquire different amounts of $x$-$p_x$ correlation, 
see plots B and C in Fig.~\ref{fig:emittance_evolution}, 
and the plot for $p_1$ in Fig.~\ref{fig:slices_fit_coeffs}.
The difference in the magnitude of the correlations for the tail and head slices
(in other words, the tilt angles of the slices in the phase space)
is due to the acquired momentum $p_r \propto E_r^\mathrm{lin}/v_z$ that 
changes (linearly as we shall see) along the bunch because of both $E_r^\mathrm{lin}$ and $v_z$. 
The space-charge field of the egg-like bunch Fig.~\ref{fig:emittance_evolution} 
can be calculated in the closed analytical form but unfortunately it is too cumbersome.
Therefore, in Fig.~\ref{fig:space-charge_field} we show a graphical illustration of
how $E_r^\mathrm{lin}$ changes \textit{linearly} in the bunch. As a result,
the tail slice experiences stronger defocusing 
than the head one. In addition, there is almost a linear energy chirp 
in the bunch.    

The difference in defocusing strength leads 
to a relative rotation between the slices in the $x-p_x$ plane.
Simultaneously, as the bunch dynamics evolves further, 
the egg-like shape transforms into a proper ellipsoidal shape (not shown in plots)
because the tail expands faster than the head due to tail's larger divergence.
The field $E_r^\mathrm{lin}$ becomes constant along the bunch. 
By the end of the region AC, the bunch density drops 
by 2 orders of magnitude because of bunch expansion.

Between the positions C and E, see Fig.~\ref{fig:emittance_evolution},
the bunch aspect ratio remains approximately constant whereas the bunch shape is nearly ellipsoidal.
In this region, the projected emittance reduces almost to the thermal value 
as the \textit{tail slices rotate towards the head slices} 
in the phase space due to the larger correlation in the tail. 
Specifically, consider two electrons having the same $x$-momentum $p_x$ but 
located in the tail, superscript $t$ , and in the head, superscript $h$.
After propagating a distance $l$, the separation between the electrons 
can be estimated in the ballistic approximation as
$(p_x/p_z^{h}-p_x/p_z^{t})l$. The difference is negative because of 
the head-tail energy chirp ($p_z^{h}>p_z^{t}$)
and the tail electron catches up with head one. 
This energy chirp effect leads to the self-compensation of 
the projected emittance in the region CD.

The region EG corresponds to a transition through the gun exit aperture. 
A clear and simple treatment of the radial bunch dynamics in the transition region 
can be found in the Wangler book~\cite{wangler2008rf}. 
The \textit{resulting} transverse momentum change can be estimated as $\Delta p_r = eE_\mathrm{acc}r/2\bar{v}_z$,
where $\bar{v}_z$ is the average longitudinal velocity $v_z$ in the exit aperture region. 
Because of the head-tail energy chirp ($v_z^{h}>v_z^{t}$), 
the tail electrons acquire a larger transverse kick and 
the tail slice is rotated counter-clockwise with respect to the head slice,
see the plot G in Fig.~\ref{fig:emittance_evolution}. 
The emittance peaks at the exit of the aperture. 
The emittance oscillation between the points E and E$'$ is due to the gradient of $E_\mathrm{acc}$
and is a local effect that has no impact on subsequent emittance evolution. 
We also note that the energy chirp has two components: 
(i) one due to the space-charge force, and 
(ii) the other one due to the accelerating RF field. 
The second contribution to the chirp collapses almost to zero when the bunch leaves the cavity. 

In the region GI, the tail slices catch up with the head slices 
due to a larger rotation speed proportional to $p_x/p_z$.
It is the same mechanism as in the region CE discussed in details above. 

The linear mechanism of the growth and self-compensation 
of the projected emittance can be summarised as follows:
(i) quick misalignment of the tail and head slices due to the energy chirp-dependent momentum kick 
from the space-charge force in the cathode region or from the accelerating field in the exit aperture;
(ii) slow self-alignment of the slices due to the ballistic effect. 
These two processes can be thought of as some excitation-relaxation mechanism which occurs twice:
in the gun (regions AC and CE) and in the exit aperture followed by drift space (regions E$'$G and GI) . 

To sum up, the discovered regime of emittance self-compensation is observed in a wide range of bunch charges from 160 fC to 16 pC. 
This regime is also observed for different lengths of the accelerating gun region deliberately designed 
to study possible limitations. The emittance self-compensation is, in general, robust except that 
the initial radial distribution of the bunch density must be close to a half-circular distribution
to keep space-charge forces linear. 
The effect of self-compensation in blow-out mode with strong space-charge forces seems to be universal. 
The extent and position of self-compensation depend on specific settings of the gun. 
The discovered regime allows generating bunches with the lowest possible emittance 
for a given charge in blow-out mode. As an example, we tracked the generated bunch distribution through 
a 15 MeV linear accelerator and found the emittance to be preserved on a 60-nm scale 
for 16 pC bunches. A small solenoid at the position of the emittance minimum collimates the bunch 
that is then further accelerated by a 352 MHz booster.

The authors acknowledge the Swedish Research Council (VR, project 2016-04593). 
The authors are thankful to Dr. Simone Di Mitri and Dr. H. Qian for the valuable comments. 
The cross-check of the simulation results by Dr. A.~Latina using RFtrack is greatly appreciated. 
Z.T. acknowledges the Hungarian Scientific Research Fund (OTKA) for the Grant No. 129134.

\section*{References}
\bibliography{references} 

\end{document}